\documentclass[a4paper,11pt]{book}
\usepackage{graphicx}
\usepackage[numbers,compress]{natbib}
\usepackage[latin1]{inputenc}
\usepackage{pstricks,pst-node,pst-text,pst-3d}
\usepackage{amsmath}
\usepackage{color}

\def\be{\begin{equation}}
\def\ee{\end{equation}}

\def\te{\end{equation}}
\def\bea{\begin{eqnarray}}
\def\nn{\nonumber\\}
\def\tea{\end{eqnarray}}

\title{Proceedings of the II Amazonian Symposium on Physics}

\begin{document}

\maketitle

\chapter{Analog cosmology with spinor BECs}

\author{Esteban Calzetta\footnote{calzetta@df.uba.ar}\\
\textit{Departamento de F\'\i sica, FCEyN UBA and IFIBA, CONICET, }\\
 \textit{Pabell\'on I, Ciudad Universitaria,
1428 Buenos Aires, Argentina}
}

\textbf{Abstract:} We show that the properties of spinor Bose-Einstein condensates allow us to build an analog Taub (axisymmetric Bianchi IX) Universe. We shall develop this proposal on the example of a rubidium condensate, where the relevant experiments are well within present day capabilities. A better Taub analog however would be built out of a collective Rydberg excitation.

\section{Introduction}
Analog models \cite{Unr81} are increasingly used to test fundamental issues in quantum and gravitational physics. While the quest for an experimental demonstration of the Unruh and Hawking effects dominates the field, analogs to cosmological space times are also of great interest. A large variety of physical systems have been proposed as the physical support for the analog models, from plain water \cite{WTPUL11} to truly astrophysical systems \cite{Das04}. Bose-Einstein condensates (BECs) stand out for the excellent control one may achieve  at the experimental level and the relatively thorough understanding of their physics at the theoretical one. There are several proposals to build black and white holes \cite{BFFRC08,CFRBF08,MacPar09,RPC09,Lah10,FabMay11,MFR11,MRFPBC11,ZAPS11,LRCP12} and expanding Universes \cite{Lid04,KKMTI09,LLR12} (including anisotropic Bianchi I Universes \cite{AmbWil10}) out of BECs, as well as proposals to test specific mechanisms without reproducing a complete cosmological scenario \cite{CalHu03,CalHu05,JWVG07,FSU07,RCPR08}. Following up on an earlier communication, in this contribution we shall describe a proposal to build an analog Taub Universe from a spinor BEC \cite{StaUed12}. The proposal relies on experiments which are clearly within present day capabilities, since similar ones have been already carried out.

The Bianchi IX metric describes the homogeneous vacuum solution to the classical Einstein equations with the largest number of free parameters. It has been thoroughly investigated as a likely description of the Universe close to the initial singularity. Moreover, the model displays dynamical chaos and as such it has been the principal source of our understanding of the interaction between gravity and chaos, both at the classical and quantum levels \cite{Cal12}. The Taub Universe is a particular case of the Bianchi IX one, where the evolution is axisymmetric besides homogeneous.

It is well known that both a particle with spin \cite{SCH68,SCH81} and a quantum field in a Bianchi type IX Universe \cite{Hu72,HFP73,Hu73,Hu74,SHOC85,HOC86,Cal91} are related to the quantum mechanical top. In an earlier communication we combined these insights to show that the dynamics of a cold gas of atoms with total momentum $F\neq 0$ may be used to explore the behavior of quantum matter in the Taub space time \cite{Cal08}. Indeed, a spinor Bose-Einstein condensate may be described as a field defined both in ordinary space time and in an internal space with the geometry of a sphere. A magnetic field deforms the internal sphere into an homogeneous space of the Taub class \cite{LLCF,MTW,RS75,BKL70,Mis70,Mis72,Bar82,HBC94}. 

However, in that work we did not go beyond the test field approximation. Since then, it has been realized that analog gravitational models need not be confined to the kinematics of gravitation. They may be used to explore certain aspects of space-time dynamics as well as fundamental issues such as the nature of semiclassical approximations \cite{Fis07,BFF07,GLS08,GLS09a,GLS09b,FLS11,FLS12}. With this in mind we shall reexamine our earlier proposal to show that the same effect occurs due to the magnetic field generated by the condensate itself. This will open the possibility of writing a self-consistent evolution for the model.

\section{Field theory of BEC}
For a field-theoretic description of BEC we begin with a second - 
quantized field operator $\Psi \left( x,t\right) $ which removes an
atom at the location $x$ at times $t$ \cite{PS02,CalHu08}. It obeys the canonical
commutation relations

\begin{equation}
\left[ \Psi \left( x,t\right) ,\Psi \left( y,t\right) \right] =0
\end{equation}
\begin{equation}
\left[ \Psi \left( x,t\right) ,\Psi ^{\dagger }\left( y,t\right) \right]
=\delta \left( x-y\right)  \label{etcr}
\end{equation}
The dynamics of this field is given by the Heisenberg equations of motion

\begin{equation}
-i\hbar \frac{\partial }{\partial t}\Psi =\left[ \mathbf{H},\Psi \right]
\end{equation}

\begin{equation}
\mathbf{H}=\int d^dx\;\left\{ \Psi ^{\dagger }H\Psi +V_{int}\left[\Psi ^{\dagger},\Psi\right] \right\} \label{nbodyh}
\end{equation}

\begin{equation}
H\Psi =-\frac{\hbar ^{2}}{2M}\nabla ^{2}\Psi +V_{trap}\left( x\right) \Psi
\label{sparth}
\end{equation}
The
Heisenberg equation of motion

\begin{equation}
i\hbar \frac{\partial }{\partial t}\Psi =H\Psi +\frac{\partial V_{int}}{\partial\Psi ^{\dagger }}
\label{Heisenberg}
\end{equation}
is also the classical equation of motion derived from the action

\begin{equation}
S=\int d^{d+1}x\;i\hbar \Psi ^{*}\frac{\partial }{\partial t}\Psi -\int dt\;%
\mathbf{H}  \label{action}
\end{equation}
The theory is invariant under a
global phase change of the field operator
\begin{equation}
\Psi \rightarrow e^{i\theta }\Psi ,\qquad \Psi ^{\dagger }\rightarrow
e^{-i\theta }\Psi ^{\dagger }  \label{global}
\end{equation}
The spontaneous breaking of this symmetry signals Bose-Einstein condensation, and the order parameter is identified with the wave function of the condensate \cite{Yuk11}.

\subsection{Spinor BECs}
In a spinor BEC, the  interaction of the nuclear spin $\mathbf{I}$ and the electron spin $\bf{S}$ for an electron in the ground state introduces in the Hamiltonian a hyperfine term $(A/2)\bf{F^2}$, where $\bf{F=I+S}$ and $A$ is a constant (for rubidium, $A\approx 2\;10^{-6}$eV), leading to a ground state with ${\bf F}^2=F(F+1)\neq 0$ \cite{PS02,Ho98,OhmMac98,LMSLP09}. Spinor BEC have been realized experimentally with several different species: rubidium $^{87}$Rb with $F=1$ \cite{SHLVS06,ALBSPP07,VLGS07,CHBSFZYC04} and $F=2$ \cite{KAEH04,SEKKSCABS04}; sodium $^{23}$Na with $F=1$ \cite{SISMCK98} and $F=2$ \cite{GGLLCGIPK03} and chromium $^{52}$Cr with $F=3$ \cite{LKFFMGGP07,BCZLMVKG07}. Spinor BEC can also be achieved with Rydberg atoms and molecular condensates \cite{MLLP07}. See also \cite{Lam07,UhScFi07,DamZur07a,DamZur07b,SaKaUe07a,SaKaUe07b,MiaGir07,VHLGSS07,LMFK08,LGVS08,VGLS09,MLFG09,LMKF09,BPV10}. For concreteness, we shall consier the case of rubidium \cite{Low06}. For this element, $I=3/2$ and the valence electron carries no orbital angular momentum so the atom can have either $F=1$ or $F=2$, with the former being the lowest modes, and so those that condensate.

Spinor condensates show local point interactions, spin-exchange interactions and dipole-dipole interactions. The strenght of each of these can be controlled independently, and we shall neglect the local point interactions.  

In principle we could clasify atomic states by the $z$ projection of the total angular momentum $F$ and describe theis associated destruction operators as so many independent fields \cite{Sin11,FizSac11}. However, to emphasize the symmetries linking these states, we shall seek an alternative description, describing the spinning atom as a particle in an enlarged configuration space.

The space of states of any spinning particle is a subspace of the Hilbert space of a quantum top \cite{Ros51}. The configuration space of the top is parameterized by Euler angles $\theta$, $\varphi$ and $\psi$, and the spin operators are identified with differential operators

\bea
F_x&=&\frac \hbar i \left\{\cos\left[\varphi\right]\frac{\partial}{\partial\theta}-\frac{\sin\left[\varphi\right]\cos\left[\theta\right]}{\sin\left[\theta\right]}\frac{\partial}{\partial\varphi}+\frac{\sin\left[\varphi\right]}{\sin\left[\theta\right]}\frac{\partial}{\partial\psi}\right\}\nn\\
F_y&=&\frac \hbar i \left\{\sin\left[\varphi\right]\frac{\partial}{\partial\theta}+\frac{\cos\left[\varphi\right]\cos\left[\theta\right]}{\sin\left[\theta\right]}\frac{\partial}{\partial\varphi}-\frac{\cos\left[\varphi\right]}{\sin\left[\theta\right]}\frac{\partial}{\partial\psi}\right\}\nn\\
F_z&=&\frac \hbar i\frac{\partial}{\partial\varphi}
\tea
L. S. Schulman used this to develop a path integral for spin \cite{SCH68}

Because of the hyperfine splitting, the Hamiltonian $H$ for a spinor condensate contains a term $(A/2)\bf{F}^2$ (for rubidium, $A=2\:10^{-6}$eV). After second quantization, this term becomes ($\chi^a=\left(\theta, \varphi,\psi\right)$)

\be
H_0=\frac{A}{2}\int d^3\chi\;\sqrt{g_0}\;\:g_0^{ab}\frac{\partial\Psi^{\dagger}}{\partial\chi^a}\frac{\partial\Psi}{\partial\chi^b}
\label{hache0}
\ee
where  $g_{0ab}$ is the metric of the sphere, given by

\be
ds_0^2=d\theta^2+d\varphi^2+d\psi^2+2\cos\theta\:d\varphi d\psi
\label{0metric}
\ee

\section{Self-consistent magnetic field}

Since all retardation effects are negligible, to couple a magnetic field to the condensate we add new terms to the Hamiltonian

\be
\int d^{d}x\!\left\{\frac1{2\mu_0}\vec{B}^2-\frac1{2\alpha}\left(\vec{\nabla}\cdot \vec{A}\right)^2-\vec{B}\vec{M}\right\}
\label{Maxwell}
\ee
where $\vec{B}=\vec{\nabla}\times\vec{A}$ and $\vec{M}$ is the magnetization; $\mu_0$ is the vacuum permeability and $\alpha$ the gauge fixing parameter. 

The point is that the nuclear and electron spin contribute in a different way to the magnetization

\be
\vec{M}=\frac1{\hbar}\left(\mu_N\vec{I}+\mu_e\vec{S}\right)\approx\frac1{\hbar}\mu_e\vec{S}
\label{gyro}
\ee
and since $\left[F^2,\vec{S}\right]\neq 0$, it mixes $F=1$ and $F=2$ states. Observe that $\mu_e=-2\mu_B$, where $\mu_B$ is Bohr's magneton $e/2m_e$, the factor of $2$ comes from the electron's g-factor, and the sign because the electron is negatively charged.

In general, the matrix elements of the magnetization may be computed from the Wigner-Eckart theorem \cite{Edm57}. However, in this simple case a direct evaluation is easiest. Let us adopt a local frame such that $\vec{B}$ points in the $z$ direction. Then we only need the matrix elements of $S_z$, which moreover commutes with $F_z$. Obviously, these matrix elements are easiest to compute in states $\left|I_z=m_I,S_z=m_z\right\rangle$ where the nuclear and electron spin are well defined; we need them for states $\left|F,F_z\right\rangle$ where total angular momentum and its $z$ projection are well defined.

Since

\be
\left|2,2\right\rangle =\left|I_z=3/2,S_z=1/2\right\rangle
\label{22}
\te
we have

\be
\frac1{\hbar}S_z\left|2,2\right\rangle =1/2\left|2,2\right\rangle
\te
Applying $F_-$ to both sides of eq. (\ref{22}) we get

\be
\left|2,1\right\rangle =\frac{\sqrt{3}}2\left|I_z=1/2,S_z=1/2\right\rangle+\frac{1}2\left|I_z=3/2,S_z=-1/2\right\rangle
\label{21}
\te
The other linearly independent combination is

\be
\left|1,1\right\rangle =\frac{-1}2\left|I_z=1/2,S_z=1/2\right\rangle+\frac{\sqrt{3}}2\left|I_z=3/2,S_z=-1/2\right\rangle
\label{11}
\te
Therefore

\bea
\frac1{\hbar}S_z\left|2,1\right\rangle &=&\frac{\sqrt{3}}4\left|I_z=1/2,S_z=1/2\right\rangle-\frac{1}4\left|I_z=3/2,S_z=-1/2\right\rangle\nn
&=&\frac{1}4\left|2,1\right\rangle-\frac{\sqrt{3}}4\left|1,1\right\rangle
\label{sz21}
\tea

\be
\frac1{\hbar}S_z\left|1,1\right\rangle =-\frac{\sqrt{3}}4\left|2,1\right\rangle-\frac{1}4\left|1,1\right\rangle
\label{sz11}
\te
Applying $F_-$ to eqs. (\ref{21}) and (\ref{11}) we get

\bea
\left|20\right\rangle&=&\frac1{\sqrt{2}}\left(\left|I_z=-1/2,S_z=1/2\right\rangle+\left|I_z=1/2,S_z=-1/2\right\rangle\right)\nn
\left|10\right\rangle&=&\frac1{\sqrt{2}}\left(-\left|I_z=-1/2,S_z=1/2\right\rangle+\left|I_z=1/2,S_z=-1/2\right\rangle\right)
\label{2010}
\tea
Therefore

\bea
\frac1{\hbar}S_z\left|2,0\right\rangle &=&\frac{-1}2\left|10\right\rangle\nn
\frac1{\hbar}S_z\left|1,0\right\rangle &=&\frac{-1}2\left|20\right\rangle
\label{sz2010}
\tea
Finally, by symmetry 

\bea
\frac1{\hbar}S_z\left|2,-1\right\rangle &=&\frac{-1}4\left|2,-1\right\rangle+\frac{\sqrt{3}}4\left|1,-1\right\rangle\nn
\frac1{\hbar}S_z\left|1,-1\right\rangle &=&\frac{\sqrt{3}}4\left|2,-1\right\rangle+\frac{1}4\left|1,-1\right\rangle
\label{sz2-11-1}
\tea

\be
\frac1{\hbar}S_z\left|2,-2\right\rangle =-1/2\left|2,-2\right\rangle
\te

\section{Low energy effective Hamiltonian}
We are now ready to obtain the dynamics of the low energy $F=1$ modes by explicitly integrating out the energetic $F=2$ modes. If moreover we neglect the kinetic energy of the $F=2$ atoms, then this integration can be carried out independently at each point in the trap. We shall consider only one such point, where moreover we assume $\vec{B}=B\hat{z}$. Let us expand

\be
\Psi=\sum_{F=1,2}\sum_{M=-F}^{F}\sum_{K=-F}^{F}f_{FMK}\left(t\right)Y_{FMK}\left(\theta,\varphi,\psi\right)
\te
where the modes $Y_{FMK}$ are common eigenfunctions of $F^2$, $F_z$ and $F_{\xi}=\left(\hbar /i\right)\left(\partial /\partial\psi\right)$ with eigenvalues $\hbar^2F\left(F+1\right)$, $\hbar M$ and $\hbar K$ respectively.

We shall consider only the one-loop approximation whereby the $F=2$ modes are considered as free except for their coupling to the $F=1$ modes \cite{CalHu08}. Within this approximation, the $F=2$, $M=\pm 2$ and the $F=2$, $K=\pm 2$ modes may be ignored. For the other modes we obtain the Hamiltonian

\be
H_2=\sum_{M=-1}^1\sum_{K=-1}^1\left\{\left[\hbar\omega_2-M\chi_2\right]f_{2MK}^{\dagger}f_{2MK}+f_{2MK}^{\dagger}J_{1MK}
+J_{1MK}^{\dagger}f_{2MK}\right\}
\te
where

\bea
\omega_2&=&\frac{3A}{\hbar}\nn
\chi_2&=&\frac{\mu_eB}4\nn
J_{1MK}&=&\frac{\mu_eB}2\left[1+\frac{\sqrt{3}}2M-M^2\right]f_{1MK}
\tea
Integrating out the $F=2$ modes within the IN-OUT formalism \cite{CalHu08} yields a non-local effective Hamiltonian

\be
H_{eff}=-\frac{i}{\hbar}\sum_{M=-1}^1\sum_{K=-1}^1\int_{t'>t}\!dt'\;\left\langle T\left[f_{2MK}\left(t'\right)f_{2MK}^{\dagger}\left(t\right)\right]\right\rangle J_{1MK}^{\dagger}\left(t'\right)J_{1MK}
\te
Where $T$ stands for time ordering. Keeping only the lowest order term in $B$ and assuming the $F=2$ modes are in their vacuum state we get

\be
\left\langle T\left[f_{2MK}\left(t'\right)f_{2MK}^{\dagger}\left(t\right)\right]\right\rangle=e^{-i\omega_2\left(t'-t\right)}
\theta\left(t'-t\right)
\te
To obtain a local effective Hamiltonian we further approximate

\be
J_{1MK}^{\dagger}\left(t'\right)=e^{i\omega_1\left(t'-t\right)}J_{1MK}^{\dagger}\left(t\right)
\te
where $\omega_1=A/\hbar$. Then the $t'$ integral may be evaluated and we get

\be
H_{eff}=\frac{\mu_B^2B^2}{8A}\sum_{M=-1}^1\sum_{K=-1}^1\left[M^2-4\right]f_{1MK}^{\dagger}f_{1MK}
\te

\section{The effective dynamics as a self-consistent metric}
To be able to interpret the effective dynamics of the $F=1$ modes as the evolution of a field in a nontrivial space time, we write again the Hamiltonian as an integral over Euler angles

\be
H_{eff}=\frac{\mu_B^2B^2}{8A}\int d^3\chi\;\sqrt{g_0}\;\:\left[\frac{\partial\Psi^{\dagger}}{\partial\varphi}\frac{\partial\Psi}{\partial\varphi}-4\Psi^{\dagger}\Psi\right]
\label{hache01}
\ee
where only the $F=1$ components of $\Psi$ are retained. 
Comparing to the case where only the hyperfine term is present, we see that the metric of the sphere has been deformed into

\be
ds_B^2=\left[1+{\cal B}^2\right]d\theta^2+d\varphi^2+\left[1+{\cal B}^2\sin^2\theta\right]d\psi^2+2\cos\theta\:d\varphi d\psi
\label{Bmetric}
\te
where

\be
{\cal B}=\frac{\mu_BB}{2A}
\label{calB}
\ee
To see that this is indeed a Taub metric, identify

\bea
\omega_1&=&d\varphi +\cos\theta\;d\psi\nn\\
\omega_2&=&\cos\varphi\,d\theta +\sin\varphi\sin\theta\;d\psi\nn\\
\omega_3&=&-\sin\varphi\,d\theta +\cos\varphi\sin\theta\;d\psi
\label{forms}
\tea
The metric can then be written in terms of Misner parameters as

\be
ds_B^2=e^{2\Omega}\left\{e^{-2\beta_+}\omega_1^2+e^{\beta_++\sqrt{3}\beta_-}\omega_2^2+e^{\beta_+-\sqrt{3}\beta_-}\omega_3^2\right\}
\label{Charlie2}
\ee
where $\beta_-=0$, $\beta_+=\Omega$ and

\be
\Omega =\frac{1}{3}\ln\left[1+{\cal B}^2\right]
\label{Omega}
\ee
Physically, the physical degree of freedom in the metric is the same as in the magnetic field, and so there is no ``Einstein'' equation independent of Maxwell's equations. Moreover, this equation is not local in the internal space, rather the magnetic field is coupled to two global quantities. The dynamic equation is the Ampere Law $\nabla\times\vec{H}=0$, where $\vec{H}$ is defined as the variational derivative of the total effective Hamiltonian with respect to $\vec{B}$. Therefore we get

\be
{H}^i=\frac1{\mu_0}{B}^i-{M}^i-c_0{B}^i+C^i_j{B}^j
\label{Ampere2}
\ee

\be
c_0=\frac{\mu_B^2}{A}\int d^3\chi\;\sqrt{g_0}\;\Psi^{\dagger}\Psi
\label{hache1}
\ee

\be
C^i_j=\frac{\mu_B^2}{8A\hbar^2}\int d^3\chi\;\sqrt{g_0}\;\Psi^{\dagger}\left\{\hat{F}^i,\hat{F}_j\right\}\Psi
\label{hache2}
\ee
Although very different in form to Einstein dynamics, this equation could be used, for example, to explore the validity of the semiclassical approximation in this problem.

\section{Final remarks}

 We have shown that a spinor BEC with dipolar interactions has a natural representation as a field on an internal space whose scale factor and shape depend upon a self-consistent magnetic field. In this sense, there is a well defined sense in which the ``field'' backreacts on the magnetic field and we have a self consistent dynamics coupling the matter field and the Maxwell field. This can be used to explore both the quantum dynamics of internal spaces with nontrivial geometry and fundamental issues in semiclassical theories.

One important difference between this problem and the full field theoretic ones is that only a handful of modes really represent physical configurations.  The problem becomes more ``field like'' the larger the number of modes. This can be accomplished, for example, by considering the dipole interaction of Rydberg collective excitations \cite{Gal94,Hei08,Hei08b,HNP10,LHL12}. Such excitations have already been built in the laboratory in $D$ states with very large values of $n$ ($n>50$) \cite{Vit11,Vit12}, and therefore couple to a much larger number of modes.

The work presented in this communication can be extended in several ways. In a dynamical setting, we expect the $F=2$ modes we have integrated out will behave as an environment for the relevant $F=1$ modes. Therefore, besides the effective field we have studied, we expect there will arise dissipation, noise and decoherence effects \cite{CalHu97,CalHu08}. These are major considerations for the assesment of current proposals for building analog models in the laboratory \cite{Wue08,LomTur12}.

\section*{Acknowledgments}

We warmly thank the organizers for the invitation to be part of the ``II Amazonian Symposium on Physics - Analogue Models of Gravity 30 Years Celebration'' where this work was presented.

This work has been supported by CONICET, UBA and  ANPCyT (Argentina)

\end{document}